\documentclass[a4paper,12pt]{article}
\usepackage{amssymb}

\def\hA{\hat{A}}
\def\hl{\hat{\lambda}}
\def\hF{\hat{F}}
\def\hD{\hat{D}}
\def\dt{\delta\theta}
\def\mfrac#1#2{\hbox{\large ${#1\over #2}$}}
\begin{document}
\begin{titlepage}
\title{
\hfill\parbox{4cm}
{\normalsize KUNS-1606\\{\tt hep-th/9909139}}\\
\vspace{1cm}
Comments on Gauge Equivalence\\ in Noncommutative Geometry}
\author{
Tsuguhiko {\sc Asakawa}\thanks{{\tt
    asakawa@gauge.scphys.kyoto-u.ac.jp}}
and
Isao {\sc Kishimoto}\thanks{{\tt
    ikishimo@gauge.scphys.kyoto-u.ac.jp}}
\\[7pt]
{\it Department of Physics, Kyoto University, Kyoto 606-8502, Japan}
}
\date{\normalsize September, 1999}
\maketitle
\thispagestyle{empty}

\begin{abstract}
\normalsize
We investigate the transformation from ordinary gauge field to
noncommutative one which was introduced by N.~Seiberg and E.~Witten
(hep-th/9908142). It is shown that the general transformation which is
determined only by gauge equivalence has a path dependence in
`$\theta$-space'. This ambiguity is negligible when we compare
the ordinary Dirac-Born-Infeld action with the noncommutative one in
the $U(1)$ case, because of the $U(1)$ nature and slowly varying
field approximation. However, in general, in the higher derivative
approximation or in the $U(N)$ case, the ambiguity cannot be neglected
due to its noncommutative structure. This ambiguity corresponds to the
degrees of freedom of field redefinition.
\end{abstract}

\end{titlepage}

\section{Introduction}

Gauge theories on noncommutative spaces have been investigated for
many years from mathematical and physical viewpoint (\cite{Connes} and 
the references in \cite{SW}). 
Especially in string theory, 
the worldvolume theory of 
D-branes in a background B-field is described by noncommutative
Yang-Mills or Dirac-Born-Infeld theory. 

Recently, Seiberg and Witten~\cite{SW} argued the equivalence between ordinary
gauge field theory and the noncommutative one as the low energy effective
theories of open strings: they arise from the same two-dimensional field
theory regularlized in different ways, so that there must be a
transformation among them. In \cite{SW}, this transformation is
uniquely given by the gauge equivalence relation, and this implies the 
equivalence between ordinary Dirac-Born-Infeld action and the
noncommutative one. 

In this short note, we re-examine the validity of above arguments and 
point out that the transformation of \cite{SW} has in general ambiguities. 
In section 2, we begin with the gauge equivalence relation between two
nearby points in the `$\theta$-space'
and show that there is ambiguity with arbitrary constant
parameters.
Then, we discuss the path dependence of the transformation in
the `$\theta$-space', which is found by considering the commutator of
two transformations. 
This implies the existence of another ambiguity.
In section 3 we investigate these ambiguities from different viewpoint.
Next in section 4, we consider the $U(1)$ case in slowly varying field
approximation. This is the situation of \cite{SW} in comparing the
ordinary Dirac-Born-Infeld action with the noncommutative one. In this
case the ambiguities do not affect the result of \cite{SW},
because of the $U(1)$ nature and of neglecting the higher derivative terms. 
In section 5, we summarize the paper and give some discussions.
In ``Note Added'', we argue that the path dependence can be reduced to 
the field redefinition.

\section{Gauge Equivalence Relation}
\label{path}
In \cite{SW}, they obtained  a transformation from ordinary gauge field
$A_{i}$ (and gauge parameter $\lambda$) to noncommutative gauge
field $\hA_{i}$ (and gauge parameter $\hl$) by demanding the gauge
equivalence relation between them. However, we show that
their statement has generally ambiguities. 
Here we investigate the gauge equivalence relation carefully.

Consider a noncommutative, associative algebra denoted by ${\cal
 A}_{\theta}=({\frak g}\otimes C^{\infty},*)$, where ${\frak g}$ is some
 Lie algebra and the $*$ product is defined to be the tensor product
 of matrix multiplication with the product of functions such as
$f(x)*g(x):=\exp\left(i{\theta^{kl}\over 2}{\partial\over\partial
 y^k}{\partial\over\partial z^l}\right)f(y)g(z)|_{y=z=x}$ with
 constant antisymmetric tensor $\theta^{kl}=-\theta^{lk}$.
Note that $\theta$'s are arbitrary parameters. We denote this
 parameter space of the whole set of algebra $\{{\cal
 A}_{\theta}\}_{\theta\in\vartheta}$ as $\theta$-space $\vartheta$.
 
We assume there exists some sort of mapping from ${\cal A}_{\theta}$
to another ${\cal A}_{\tilde\theta}$ in a way that preserve gauge
equivalence relation which is described by the following equation in terms of
gauge fields and gauge parameters ${\tilde A}_{i}(\hA),{\tilde\lambda}(\hA,\hl)\in{\cal A}_{\tilde\theta}$ and $\hA_{i},\hl\in{\cal A}_{\theta}$:
\begin{equation}
{\tilde A}_{i}(\hA)+{\tilde\delta}_{\tilde\lambda}{\tilde A}_{i}(\hA)={\tilde A}_{i}(\hA+{\hat\delta}_{\hl}\hA),
\label{eq:gaugeeqv}
\end{equation}
where ${\hat\delta}_{\hl}$ is the gauge transformation with infinitesimal $\hl$, i.e., 
${\hat\delta}_{\hl}\hA_{i}=\hD_{i}\hl=\partial_{i}\hl-i[\hA_{i},\hl].$
\footnote{In this paper, $[\hA,{\hat B}]=\hA*{\hat B}-{\hat B}*\hA, \
  \{\hA,{\hat B}\}=\hA*{\hat B}+{\hat B}*\hA$.}
and likewise for ${\tilde\delta}_{\tilde\lambda}$. This relation means
that the diagram below is commutative.

\begin{center}
%WinTpicVersion2.16
\unitlength 0.1in
\begin{picture}(23.60,13.05)(22.90,-17.35)
% STR 2 0 3 0
% 3 2510 900 2510 1000 2 0
% $\hA_i'$
\put(25.1000,-6.0000){\makebox(0,0)[lb]{$\hA_i'$}}%
% STR 2 0 3 0
% 3 2520 2120 2520 2220 2 0
% $\hA_i$
\put(25.2000,-18.2000){\makebox(0,0)[lb]{$\hA_i$}}%
% STR 2 0 3 0
% 3 4410 2110 4410 2210 2 0
% ${\tilde A}_i$
\put(44.1000,-18.1000){\makebox(0,0)[lb]{${\tilde A}_i$}}%
% STR 2 0 3 0
% 3 4390 900 4390 1000 2 0
% ${\tilde A}_i'$
\put(43.9000,-6.0000){\makebox(0,0)[lb]{${\tilde A}_i'$}}%
% STR 2 0 3 0
% 3 2290 1560 2290 1660 2 0
% ${\hat\delta}_{\hl}$
\put(22.9000,-12.6000){\makebox(0,0)[lb]{${\hat\delta}_{\hl}$}}%
% STR 2 0 3 0
% 3 4650 1570 4650 1670 2 0
% ${\tilde\delta}_{\tilde\lambda}$
\put(46.5000,-12.7000){\makebox(0,0)[lb]{${\tilde\delta}_{\tilde\lambda}$}}%
% VECTOR 2 0 3 0
% 2 2950 930 4230 930
% 
\special{pn 8}%
\special{pa 2950 530}%
\special{pa 4230 530}%
\special{fp}%
\special{sh 1}%
\special{pa 4230 530}%
\special{pa 4163 510}%
\special{pa 4177 530}%
\special{pa 4163 550}%
\special{pa 4230 530}%
\special{fp}%
% VECTOR 2 0 3 0
% 2 4520 2000 4520 1140
% 
\special{pn 8}%
\special{pa 4520 1600}%
\special{pa 4520 740}%
\special{fp}%
\special{sh 1}%
\special{pa 4520 740}%
\special{pa 4500 807}%
\special{pa 4520 793}%
\special{pa 4540 807}%
\special{pa 4520 740}%
\special{fp}%
% CIRCLE 2 0 3 0
% 4 3590 1530 3950 1690 3770 1880 3220 1700
% 
\special{pn 8}%
\special{ar 3590 1130 394 394  2.7109001 6.2831853}%
\special{ar 3590 1130 394 394  0.0000000 1.0957856}%
% SARROW 2 0 3 1
% 2 3783 1873 3770 1880
% 
\special{pn 8}%
\special{pa 3783 1473}%
\special{pa 3770 1480}%
\special{fp}%
\special{sh 1}%
\special{pa 3770 1480}%
\special{pa 3838 1466}%
\special{pa 3817 1455}%
\special{pa 3819 1431}%
\special{pa 3770 1480}%
\special{fp}%
% VECTOR 2 0 3 0
% 2 2950 2130 4230 2130
% 
\special{pn 8}%
\special{pa 2950 1730}%
\special{pa 4230 1730}%
\special{fp}%
\special{sh 1}%
\special{pa 4230 1730}%
\special{pa 4163 1710}%
\special{pa 4177 1730}%
\special{pa 4163 1750}%
\special{pa 4230 1730}%
\special{fp}%
% VECTOR 2 0 3 0
% 2 2610 2010 2600 1100
% 
\special{pn 8}%
\special{pa 2610 1610}%
\special{pa 2600 700}%
\special{fp}%
\special{sh 1}%
\special{pa 2600 700}%
\special{pa 2581 767}%
\special{pa 2601 753}%
\special{pa 2621 766}%
\special{pa 2600 700}%
\special{fp}%
\end{picture}%
\end{center}

Especially in the case of nearby points in $\vartheta$,
i.e., $\tilde\theta=\theta+\dt$ with infinitesimal $\dt$,
 eq.(\ref{eq:gaugeeqv}) is
written in the variational form  as
\begin{equation}
{\hat\delta}_{\hl}\delta\hA_{i}=\delta{\hat\delta}_{\hl}\hA_{i}
\label{eq:gaugeeqvinf}
\end{equation}
by writing ${\tilde A}=\hA+\delta\hA(\hA)+{\cal O}(\delta\theta^2)$, 
${\tilde \lambda}=\hl+\delta\hl(\hA, \hl)+{\cal O}(\delta\theta^2)$ 
and expanding (\ref{eq:gaugeeqv}) to the first order in $\delta\theta$. 

We first look for the solution of (\ref{eq:gaugeeqv}) by using the
method described in the next section. Eq.(\ref{eq:gaugeeqv}) can be
easily rewritten as\footnote{Use following relations
\begin{equation}
\delta\{f,g\}=\{\delta f,g\}+\{f,\delta g\}+{i\over
  2}\delta\theta^{pq}[\partial_{p}f,\partial_{q}g],\ 
\delta\hD_{i}f=\hD_{i}\delta f-i[\delta\hA_{i},f]+{i\over
  2}\delta\theta^{pq}\{\partial_{p}\hA_{i},\partial_{q}f\}.\nonumber
\end{equation}}
\begin{equation}
{\hat\delta}_{\hl}\delta\hA_{i}-\hD_{i}\delta\hl+i[\delta\hA_{i},\hl]=-\mfrac12\dt^{kl}\{\partial_{k}\hA_{i},\partial_{l}\hl\},\nonumber
\label{eq:O1}
\end{equation}
which corresponds to the $n=1$ case of (\ref{eq:On}).
Note that this form is actually the same one
as given in \cite{SW}: the $\delta\theta$ version of eq.(3.4). 
It is solved most generally by (see next section for detail)
\begin{eqnarray}
\label{eq:delta}
\delta \hA_{i}&=&-\mfrac14\delta\theta^{kl}\{\hA_k,\partial_l\hA_{i}+\hF_{li}\}+\alpha\delta\theta^{kl}\hD_{i}\hF_{kl}+\beta\delta\theta^{kl}\hD_{i}[\hA_k,\hA_l],\nonumber\\
\delta\hl&=&\mfrac14\delta\theta^{kl}\{\partial_k\hl,\hA_l\}+2\beta\delta\theta^{kl}[\partial_k\hl,\hA_l],\nonumber\\
\delta\hF_{ij}&=&\mfrac14\delta\theta^{kl}\left( 2\{\hF_{i k},\hF_{j l}\}-\{\hA_k,\hD_l\hF_{ij}+\partial_l\hF_{ij}\}\right)\nonumber\\
& &-i\alpha\delta\theta^{kl}[\hF_{ij},\hF_{kl}]-i\beta\delta\theta^{kl}[\hF_{ij},[\hA_k,\hA_l]],
\end{eqnarray}
where
$\hF_{ij}=\partial_i\hA_j-\partial_j\hA_i-i[\hA_i,\hA_j]$
is the field strength and $\alpha,\beta$ are arbitrary
 constants. ($\alpha=\beta=0$ case corresponds to (3.8) of \cite{SW}.)
The presence of arbitrary parameters $\alpha$ and $\beta$ implies
that, with the requirement of the gauge equivalence alone, there 
exists in general ambiguity in determining an infinitesimal
mapping. However, note that this ambiguity has rather trivial origin because 
we look for two functions $\delta\hA_{i}$, $\delta\hl$ as the solution 
of one equation (\ref{eq:gaugeeqvinf}), and that the terms with 
$\alpha,\beta$ have formally a form of 
some gauge transformation. Recall that the mapping that satisfies
the gauge equivalence relation is the one which maps gauge orbits from
${\cal A}_{\theta}$ to ${\cal A}_{\tilde\theta}$ rather than
gauge fields themselves. Therefore, this kind of ambiguity is not
relevant when we discuss only gauge equivalence classes. 

However, applying $\delta\theta$-variation twice, we will encounter
the second kind of ambiguities. Denote each variation as $\delta_1$
and $\delta_2$, respectively, which are in general different direction 
with each other in the $\theta$-space, and consider their
commutation relation acting on $\hA_{i}$:
\begin{equation}
[\delta_{1},\delta_{2}]\hA_{i},
\label{eq:pathdep}
\end{equation}
which measures the `path dependence' in $\theta$-space
$\vartheta$. Using the transformation (\ref{eq:delta}) twice, we
obtain explicitly  
\begin{eqnarray}
\label{eq:del12A}
&&[\delta_{1},\delta_{2}]\hA_{i}\nonumber\\
&=&\delta_{1}\left(-\mfrac14\delta\theta^{kl}_{2}\{\hA_{k},\partial_{l}\hA_{i}+\hF_{li}\}+\alpha\delta\theta^{kl}_{2}\hD_{i}\hF_{kl}+\beta\delta\theta^{kl}_{2}\hD_{i}[\hA_k,\hA_l]\right)-(1\leftrightarrow 2)\nonumber\\
&=&{1\over 16}\delta\theta^{kl}_{2}\delta\theta^{pq}_{1}\biggl(4i[\hF_{kp},\partial_{l}\partial_{q}\hA_{i}]
+4[\hF_{kp},[\hA_{l},\partial_{q}\hA_{i}]+[\hA_{q},\partial_{l}\hA_{i}]]\nonumber\\
&&-[\partial_{k}\hA_{p}+\hF_{kp},[\hA_{l},\hD_{i}\hA_{q}]]+[\partial_{p}\hA_{k}+\hF_{pk},[\hA_{q},\hD_{i}\hA_{l}]]\nonumber\\
&&+\{\hA_{k},\{\hF_{lq},\hD_{i}\hA_{p}\}\}-\{\hA_{p},\{\hF_{ql},\hD_{i}\hA_{k}\}\}\nonumber\\
&&-i\{\hA_{p},\{\hA_{k},[\hA_{l},\hD_{i}\hA_{q}]\}\}+i\{\hA_{k},\{\hA_{p},[\hA_{q},\hD_{i}\hA_{l}]\}\}\nonumber\\
&&+2i[\partial_{p}\hA_{k},\hD_{i}\partial_{q}\hA_{l}]-2i[\partial_{k}\hA_{p},\hD_{i}\partial_{l}\hA_{q}]\nonumber\\
&&-[[\hA_{k},\hA_{p}],\hD_{i}\partial_{q}\hA_{l}]+[[\hA_{p},\hA_{k}],\hD_{i}\partial_{l}\hA_{q}]\nonumber\\
&&-\{\hA_{k},\{\hA_{p},\hD_{i}\partial_{q}\hA_{l}\}\}+\{\hA_{p},\{\hA_{k},\hD_{i}\partial_{l}\hA_{q}\}\}\biggr)\nonumber\\
&&+\hD_i\biggr(\dt_2^{kl}\dt_1^{pq}\Big(i\alpha^2[\hF_{kl},\hF_{pq}]+i\beta^2[[\hA_k,\hA_l],[\hA_p,\hA_q]]\nonumber\\
&&+i\alpha\beta([[\hA_k,\hA_l],\hF_{pq}]-[[\hA_p,\hA_q],\hF_{kl}])\nonumber\\
&&+\mfrac14\alpha(\{\partial_k\hF_{pq},\hA_l\}-\{\partial_p\hF_{kl},\hA_q\})\nonumber\\
&&+\mfrac14\beta(\{\partial_k[\hA_p,\hA_q],\hA_l\}-\{\partial_p[\hA_k,\hA_l],\hA_q\})\Big)\nonumber\\
&&+\delta\theta_2^{kl}\delta_1(\alpha\hF_{kl}+\beta[\hA_k,\hA_l])
-\delta\theta_1^{pq}\delta_2(\alpha\hF_{pq}+\beta[\hA_p,\hA_q])\biggr).
\end{eqnarray}
Note that the sum of all $\alpha,\beta$ dependent terms again has the form
of some gauge transformation (with $\hA_i$ dependent parameter).
This is easily understood by noticing that the gauge
transformations are closed under commutation relations and the
requirement (\ref{eq:gaugeeqvinf}). 
Contrary, the $\alpha,\beta$ independent terms are
 nontrivial and they do not vanish in general. That is,  
there exists path dependence if we repeat variations
more than one step in $\delta\theta$. In terms of the gauge equivalence,
(\ref{eq:del12A}) means the following. In the same sense as we
discussed below (\ref{eq:delta}) for the one-step variation, a gauge
orbit in ${\cal  A}_{\theta}$ is mapped to an orbit in ${\cal
  A}_{\theta+\delta\theta_1+\delta\theta_2}$, but now depending on the
path: orbits mapped along two paths are not the same. 

\begin{center}
%WinTpicVersion2.16
\unitlength 0.1in
\begin{picture}(33.00,25.80)(13.60,-27.80)
% LINE 2 0 3 0
% 4 2800 610 2810 2780 2080 1000 2080 3160
% 
\special{pn 8}%
\special{pa 2800 210}%
\special{pa 2810 2380}%
\special{fp}%
\special{pa 2080 600}%
\special{pa 2080 2760}%
\special{fp}%
% LINE 2 0 3 0
% 4 3840 600 3840 2790 3140 970 3140 3180
% 
\special{pn 8}%
\special{pa 3840 200}%
\special{pa 3840 2390}%
\special{fp}%
\special{pa 3140 570}%
\special{pa 3140 2780}%
\special{fp}%
% LINE 2 0 3 0
% 2 3080 980 3090 3180
% 
\special{pn 8}%
\special{pa 3080 580}%
\special{pa 3090 2780}%
\special{fp}%
% LINE 2 0 3 0
% 6 3290 1540 3380 1600 3380 1610 3270 1690 2470 1770 2430 1930
% 
\special{pn 8}%
\special{pa 3290 1140}%
\special{pa 3380 1200}%
\special{fp}%
\special{pa 3380 1210}%
\special{pa 3270 1290}%
\special{fp}%
\special{pa 2470 1370}%
\special{pa 2430 1530}%
\special{fp}%
% LINE 2 0 3 0
% 2 2450 1920 2570 1930
% 
\special{pn 8}%
\special{pa 2450 1520}%
\special{pa 2570 1530}%
\special{fp}%
% LINE 2 0 3 0
% 2 2570 2110 2760 2200
% 
\special{pn 8}%
\special{pa 2570 1710}%
\special{pa 2760 1800}%
\special{fp}%
% LINE 2 0 3 0
% 2 2750 2200 2570 2310
% 
\special{pn 8}%
\special{pa 2750 1800}%
\special{pa 2570 1910}%
\special{fp}%
% LINE 2 0 3 0
% 2 3450 1790 3410 1950
% 
\special{pn 8}%
\special{pa 3450 1390}%
\special{pa 3410 1550}%
\special{fp}%
% LINE 2 0 3 0
% 2 3420 1950 3540 1950
% 
\special{pn 8}%
\special{pa 3420 1550}%
\special{pa 3540 1550}%
\special{fp}%
% STR 2 0 3 0
% 3 3220 1360 3220 1460 2 0
% $\dt_1$
\put(32.2000,-10.6000){\makebox(0,0)[lb]{$\dt_1$}}%
% STR 2 0 3 0
% 3 2390 2420 2390 2520 2 0
% $\dt_1$
\put(23.9000,-21.2000){\makebox(0,0)[lb]{$\dt_1$}}%
% STR 2 0 3 0
% 3 3460 2090 3460 2190 2 0
% $\dt_2$
\put(34.6000,-17.9000){\makebox(0,0)[lb]{$\dt_2$}}%
% STR 2 0 3 0
% 3 2210 1620 2210 1720 2 0
% $\dt_2$
\put(22.1000,-13.2000){\makebox(0,0)[lb]{$\dt_2$}}%
% LINE 2 0 3 0
% 2 2810 1620 3830 1610
% 
\special{pn 8}%
\special{pa 2810 1220}%
\special{pa 3830 1210}%
\special{fp}%
% LINE 2 0 3 0
% 2 3840 1610 3140 2150
% 
\special{pn 8}%
\special{pa 3840 1210}%
\special{pa 3140 1750}%
\special{fp}%
% LINE 2 0 3 0
% 2 2810 1620 2080 2180
% 
\special{pn 8}%
\special{pa 2810 1220}%
\special{pa 2080 1780}%
\special{fp}%
% LINE 2 0 3 0
% 2 2080 2190 3080 2210
% 
\special{pn 8}%
\special{pa 2080 1790}%
\special{pa 3080 1810}%
\special{fp}%
% LINE 2 2 3 0
% 4 4590 1220 4300 1420 4320 1410 4660 1400
% 
\special{pn 8}%
\special{pa 4590 820}%
\special{pa 4300 1020}%
\special{dt 0.045}%
\special{pa 4320 1010}%
\special{pa 4660 1000}%
\special{dt 0.045}%
% STR 2 0 3 0
% 3 4600 1210 4600 1310 2 0
% $\vartheta$
\put(46.0000,-9.1000){\makebox(0,0)[lb]{$\vartheta$}}%
% VECTOR 2 2 3 0
% 2 1600 2710 1590 1220
% 
\special{pn 8}%
\special{pa 1600 2310}%
\special{pa 1590 820}%
\special{dt 0.045}%
\special{sh 1}%
\special{pa 1590 820}%
\special{pa 1570 887}%
\special{pa 1590 873}%
\special{pa 1610 887}%
\special{pa 1590 820}%
\special{fp}%
% STR 2 0 3 0
% 3 1360 1910 1360 2010 2 0
% ${\hat\delta}_{\hat\lambda}$
\put(13.6000,-16.1000){\makebox(0,0)[lb]{${\hat\delta}_{\hat\lambda}$}}%
\end{picture}%

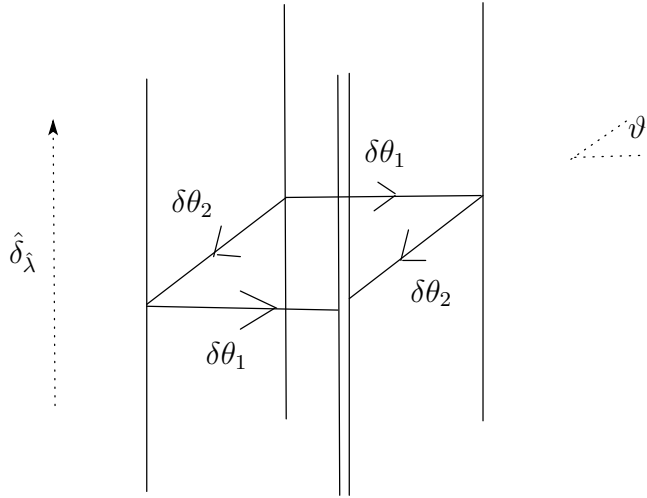
{\small A vertical line denotes gauge orbit on a point in $\theta$-space.\\
The double line denotes two different orbits on the same point.}
\end{center}

This second type of ambiguities accumulates globally in
$\theta$-space, if we consider any mapping from ${\cal A}_{\theta}$ to 
${\cal A}_{\tilde \theta}$ at a finite distance apart in
$\theta$-space.
Transformation on gauge fields is given by the integration over
$\delta\theta$ by specifying a path by hand as 
\begin{equation}
  {\tilde A}=\int_{\rm path}{\delta\hA}.
\label{eq:integral}
\end{equation}
Of course, $\delta\hA$ suffers also from the first type of
ambiguities. If we further fix $\alpha$ and $\beta$ by hand,
i.e. select a representative, then 
${\tilde A}$ is uniquely `determined'. The procedure described in
\cite{SW}, where the functional $\hA(A)$ is determined order by order
in $\theta$, is exactly the one discussed here. 
In fact, the solution of \cite{SW} corresponds to taking 
$\alpha=\beta=0$ and the `straight line' in $\theta$-space as the path
of integration.
Here the `straight line' corresponds to the formal exponentiation of
the infinitesimal transformation (\ref{eq:delta}).

Note that there exists no rule to 
select a particular path from the standpoint of gauge theory (or more
precisely a space of the whole set of 
algebra $\{{\cal A}_{\theta}\}_{\theta\in\vartheta}$.) We need some
physical requirement. In \S 4 we discuss the equivalence of
actions between ordinary gauge theory and noncommutative one in this 
point of view.

\section{More Comments on Ambiguity}
\label{more}
In this section, we investigate the gauge equivalence relation
(\ref{eq:gaugeeqv}) from another viewpoint.

To get a solution of (\ref{eq:gaugeeqv}) directly, we expand formally ${\tilde A}_i$ as the power series in $\dt={\tilde\theta}-\theta$:
\begin{equation}
{\tilde A}_{i}=\sum_{n=0}^{\infty}\hA_{i}^{(n)},\ {\tilde \lambda}=\sum_{n=0}^{\infty}\hl^{(n)},
\label{eq:expand}
\end{equation}
where $\hA_{i}^{(n)},\hl^{(n)}\in{\cal A}_{\theta}$ are of ${\cal O}(\delta\theta^n)$, and $\hA_{i}^{(0)}=\hA_{i},\hl^{(0)}=\hl$.
Substituting this formal expansion (\ref{eq:expand}) into
(\ref{eq:gaugeeqv}), the equation of ${\cal O}(\delta\theta^n)$ is 
\begin{eqnarray}
\label{eq:On}
& &{\hat\delta}_{\hl}\hA_{i}^{(n)}-\hD_{i}\hl^{(n)}+i\left[\hA_{i}^{(n)},\hl\right]\nonumber\\
&=&-i\sum\left({i\over 2}\right)^r\dt^{k_1l_1}\cdots\dt^{k_rl_r}\left[\partial_{k_1}\cdots\partial_{k_r}\hA_{i}^{(p)},\partial_{l_1}\cdots\partial_{l_r}\hl^{(q)}\right\},
\end{eqnarray}
where the summation ranges in $p+q+r=n,\ p,q,r\ge 0,\ p\ne n,\ q\ne n$,
and $[\ ,\ \}$ denotes the anti-commutator $\{\ ,\ \}$ 
(the commutator $[\ ,\ ]$) if $r$ is odd (even).
This equation implies that $\hA_{i}^{(n)},\hl^{(n)}$ on the left hand
side are determined by ${\cal O}(\dt^{n-1})$ quantities on the right
hand side.  

Concrete procedure to get $\hA_i^{(n)},\hl^{(n)}$ is as follows:
substitute the solution  $\hA_i^{(k)},\hl^{(k)}
(k=1,\dots,n-1)$  of (\ref{eq:On}) to the right hand side, express
$\hA^{(n)}, \hl^{(n)}$ as a polynomial of
$\dt^n,\hA_i,\partial_j\hA_k,\dots,\hl,\partial_l\hl,\dots$ in the most
general form with suitable indices\footnote{We assume here that a
  transformation from $\hA_i,\hl$ to ${\tilde A}_i,{\tilde\lambda}$
  can be expressed by some polynomial of $\hA_i,\hl, \partial_j\hA_k,
  \dots,\dt^{mn}$ alone and indices are contracted among them.} 
and substitute it to the left hand side of (\ref{eq:On}), then we can
determine the coefficients in the polynomial.

However, suppose there exist some functions
$\hA_{i}^{0(n)},\hl^{0(n)}$ such that
\begin{equation}
{\hat\delta}_{\hl}\hA_{i}^{0(n)}-\hD_{i}\hl^{0(n)}+i[\hA_{i}^{0(n)},\hl]=0.
\label{eq:Ol}
\end{equation}
Then $\hA_{i}^{(n)}+\hA_{i}^{0(n)},\hl^{(n)}+\hl^{0(n)}$ are also a
solution of (\ref{eq:On}) if $\hA_{i}^{(n)},\hl^{(n)}$ satisfy
(\ref{eq:On}). In fact, we can construct such
$\hA_{i}^{0(n)},\hl^{0(n)}$ as follows.

Noting that
\begin{eqnarray}
{\hat\delta}_{\hl}\hA_{i}=\partial_{i}\hl-i[\hA_{i},\hl],\ 
{\hat\delta}_{\hl}\hD_{i}\hA_{j}=\hD_{i}(\partial_{j}\hl)-i[\hD_{i}\hA_{j},\hl],\dots ,
\end{eqnarray}
and that ${\hat\delta}_{\hat\lambda}$ and the commutator $[\ ,\ ]$
satisfy Leibnitz rule, we obtain the following identity for 
any polynomial ${\hat G}$ of $\hA_{i},\hD_{i}\hA_{j},\dots$ in 
${\cal A}_{\theta}$:
\begin{eqnarray}
\label{eq:G}
{\hat\delta}_{\hl}{\hat G}(\hA_j,\hD_k\hA_l,\dots)-{\hat\delta}'_{\hl}{\hat G}(\hA_j,\hD_k\hA_l,\dots)+i[{\hat G}(\hA_j,\hD_k\hA_l,\dots),\hl]=0,
\end{eqnarray}
where ${\hat\delta}'_{\hl}$ acts like
$\partial_i\hl\cdot{\delta\over\delta\hA_i}$ i.e., replaces $\hA_i$
with $\partial_i\hl$ but does not act on $\hA_i$ in $\hD_{i}$ (hence
$\hD_i{\hat\delta}'_{\hl}={\hat\delta}'_{\hl}\hD_i$).  In the same way,
\begin{eqnarray}
{\hat\delta}_{\hl}\hF_{ij}=-i[\hF_{ij},\hl],\ {\hat\delta}_{\hl}\hD_{k}\hF_{ij}=-i[\hD_{k}\hF_{ij},\hl],\dots ,
\end{eqnarray}
lead to
\begin{equation}
\label{eq:GF}
{\hat\delta}_{\hl}{\hat
  G}^F(\hF_{jk},\hD_{l}\hF_{mn},\dots)+i\left[{\hat
    G}^F(\hF_{jk},\hD_{l}\hF_{mn},\dots),\hl\right]=0 ,
\end{equation}
where ${\hat G}^F(\hF_{jk},\hD_{l}\hF_{mn},\dots)$ is a polynomial of $\hF_{jk},\hD_{l}\hF_{mn},\dots$ in ${\cal A}_{\theta}$.

From (\ref{eq:G}) and (\ref{eq:GF}), we get one type of solution of
(\ref{eq:Ol})
\begin{eqnarray}
\hA_i^{0(n)}&=&{\hat G}^{(n)F}_{i}(\hF_{jk},\hD_{l}\hF_{mn},\dots;\dt^n)+\hD_i{\hat G}^{(n)}(\hA_j,\hD_k\hA_l,\dots;\dt^n),\nonumber\\
\hl^{0(n)}&=&{\hat\delta}'_{\hl}{\hat G}^{(n)}(\hA_j,\hD_k\hA_l,\dots;\dt^n),
\label{eq:A0n}
\end{eqnarray}
This means that there is large ambiguity due to arbitrary polynomials\\
${\hat G}^{(n)}(\hA_j,\hD_k\hA_l,\dots;\dt^n),{\hat
  G}^{(n)F}_{i}(\hF_{jk},\hD_{l}\hF_{mn},\dots;\dt^n)$ in each order
in (\ref{eq:expand}). This result is consistent with the ambiguity
due to the path dependence of \S\ref{path}.

In particular, if we take $\dt$ infinitesimal, then the ambiguity is of 
the form
\begin{eqnarray}
{\hat G}^{(1)}(\hA_j,\hD_k\hA_l,\dots;\dt^1)&=&\beta_1\dt^{kl}[\hA_k,\hA_l]+\beta_2\dt^{kl}\hD_k\hA_l,\nonumber\\
{\hat G}^{(1)F}_{i}(\hF_{jk},\hD_{l}\hF_{mn},\dots;\dt^1)&=&\alpha_1\dt^{kl}\hD_i\hF_{kl},
\label{eq:G1}
\end{eqnarray}
where $\alpha_1,\beta_1,\beta_2$ are arbitrary constants.
Substituting (\ref{eq:G1}) into (\ref{eq:A0n}), we get
\begin{eqnarray}
\hA_i^{0(1)}&=&\dt^{kl}(\alpha_1\hD_i\hF_{kl}+\beta_1\hD_i[\hA_k,\hA_l]+\beta_2\hD_i\hD_k\hA_l)\nonumber\\
&=&(\alpha_1+\mfrac12\beta_2)\dt^{kl}\hD_i\hF_{kl}+(\beta_1-\mfrac12 i\beta_2)\dt^{kl}\hD_i[\hA_k,\hA_l],\nonumber\\
\hl^{0(1)}&=&\dt^{kl}(2\beta_1[\partial_k\hl,\hA_l]+\beta_2\hD_k\partial_l\hl)
=(2\beta_1-i\beta_2)\dt^{kl}[\partial_k\hl,\hA_l].
\end{eqnarray}
By redefinition of coefficients, this is the $\alpha,\beta$ dependent term in (\ref{eq:delta}).

\section{$U(1)$ Case}
\label{u1}
In this section, we consider the case where the gauge group is $U(1)$. 
We assume here that $\hF_{ij}$ is slowly varying so that we can ignore
${\cal O}(\partial\hF)$. This approximation is adopted when we
consider the Dirac-Born-Infeld action. Precisely, we regard $\hF\sim
\partial\hA$ as ${\cal O}(1)$ and count the order by (the number of
$\partial$)$-$(that of $\hA$).
Note that
$\hD_i=\partial_i+\theta^{jk}\partial_j\hA_i\partial_k+{\cal
  O}(\partial^4\hF\partial),
\hF_{ij}=\partial_i\hA_j-\partial_j\hA_i+\theta^{kl}\partial_k\hA_i\partial_l\hA_j+{\cal
  O}(\partial^4\hF)$, and that $\alpha,\beta$ dependent terms in
$\delta\hA_i$ in (\ref{eq:delta}) are of ${\cal O}(\partial^2\hF)$,
and hence negligible.

Eq.(\ref{eq:del12A}) reduces in this approximation to
\begin{eqnarray}
\label{eq:del12u1}
[\delta_1,\delta_2]\hA_i=\mfrac14\dt_2^{kl}\dt_1^{pq}
\hD_i(\hA_k\hA_p\hF_{lq})+{\cal O}(\hA\partial^4\hF) , 
\end{eqnarray}
and in the same way, we obtain\footnote{
The first equality of (\ref{eq:del12Fu1}) is exact even in
  the $U(N)$ case and the second one is valid only in the $U(1)$ case.}
\begin{eqnarray}
[\delta_1,\delta_2]\hF_{ij}&=&
{1\over 16}\dt_2^{kl}\dt_1^{pq}\biggr(4(i[\hD_p\hF_{ik},\hD_q\hF_{jl}]-i[\hD_k\hF_{ip},\hD_l\hF_{jq}]\nonumber\\
&&+[[\hF_{ik},\hF_{jp}]+[\hF_{ip},\hF_{jk}],\hF_{lq}])\nonumber\\
&&+4(i[\hF_{kp},\partial_l\partial_q\hF_{ij}]+[\hF_{kp},[\hA_q,\partial_l\hF_{ij}]+[\hA_l,\partial_q\hF_{ij}]])\nonumber\\
&&+2i[[\hA_p,\hA_k],[\partial_l\hA_q,\hF_{ij}]+[\partial_q\hA_l,\hF_{ij}]]\nonumber\\
&&+i[\partial_q\hA_l+\hF_{ql},[\hA_p,[\hA_k,\hF_{ij}]]]-i[\partial_l\hA_q+\hF_{lq},[\hA_k,[\hA_p,\hF_{ij}]]]\nonumber\\
&&+i\{\hA_p,\{\hF_{lq},[\hA_k,\hF_{ij}]\}\}-i\{\hA_k,\{\hF_{ql},[\hA_p,\hF_{ij}]\}\}\nonumber\\
&&-\{\hA_k,\{\hA_p,[\hA_q,[\hA_l,\hF_{ij}]]\}\}+\{\hA_p,\{\hA_k,[\hA_l,[\hA_q,\hF_{ij}]]\}\}\nonumber\\
&&+i\{\hA_k,\{\hA_p,[\partial_l\hA_q,\hF_{ij}]\}\}-i\{A_p,\{\hA_k,[\partial_q\hA_l,\hF_{ij}]\}\}\nonumber\\
&&-2[\partial_p\hA_k,[\partial_q\hA_l,\hF_{ij}]]+2[\partial_k\hA_p,[\partial_l\hA_q,\hF_{ij}]]\biggr)\nonumber\\
&&+(\alpha,\beta\ {\rm dependent\ terms})\nonumber\\
&=&-\mfrac14 i\dt_2^{kl}\dt_1^{pq}[\hF_{ij},\hA_k\hA_p\hF_{lq}]+{\cal O}(\partial^4\hF).
\label{eq:del12Fu1}
\end{eqnarray}
The right hand sides of (\ref{eq:del12u1}) and (\ref{eq:del12Fu1}) have
terms that is not of ${\cal O}(\partial\hF)$  but is of the form of gauge
transformation with gauge parameter
$\mfrac14\dt_2^{kl}\dt_1^{pq}\hA_k\hA_p\hF_{lq}$.\footnote{
Notice that, if $\theta^{kl}\ne 0$, $\hF_{ij}$ is not gauge invariant
even in the $U(1)$ case.}
This means that $\hA_i$ and $\hF_{ij}$ can be determined up
to gauge transformation in such rough approximation of ignoring
${\cal  O}(\partial\hF)$. 

In \cite{SW} they showed that the ordinary Dirac-Born-Infeld
Lagrangian equals the noncommutative one up to total derivative terms
and up to ${\cal O}(\partial\hF)$.
They argued that the more general Lagrangian 
\begin{eqnarray}
\label{eq:DBIhat}
{\hat{\cal L}}_{\rm DBI}={1\over G_s}\sqrt{\det(G+\hF+\Phi)}
\end{eqnarray}
is invariant up to total derivative terms and up to ${\cal
  O}(\partial\hF)$ under the variation with respect to $\theta$.
The gauge field of ordinary Dirac-Born-Infeld theory is in ${\cal
  A}_{\theta=0}$ and noncommutative one is in ${\cal A}_{\theta\ne
  0}$.\footnote{
The antisymmetric tensor $\Phi$ is given by ${1\over
  G+\Phi}=-\theta+{1\over g+B}$, where $G,g,B$ is the open string metric,
the closed string metric and the NS 2-form field, respectively.} 
In their proof, eq.(\ref{eq:delta}) with $\alpha=\beta=0$ is used. 
There is in general ambiguity due to $\alpha,\beta$ dependence in
(\ref{eq:delta}) but this is negligible.

As we discussed in the previous sections, there is ambiguity due to
path dependence in $\theta$-space. This implies that
\begin{eqnarray}
{\hat{\cal L}}_{\rm DBI}|_{\tilde\theta}-{\hat{\cal L}}_{\rm
  DBI}|_{\theta}=\int_{\rm path}\delta{\hat{\cal L}}_{\rm DBI} .
\end{eqnarray}
However, this path dependence is in fact missing as seen from
the gauge transformation form of (\ref{eq:del12u1}) and
(\ref{eq:del12Fu1}).
Therefore their proof of equivalence between the ordinary
Dirac-Born-Infeld action and the noncommutative one (or more generally
the equivalence of the action (\ref{eq:DBIhat}) in $\theta$-space) is
also justified in our context. 
This means that in this physical input (i.e., equivalence of the
ordinary DBI action and noncommutative one) no ambiguity is
restricted.

\section{Conclusions and Discussions}
In this paper, we considered a transformation from $\hA_i,\hl\in{\cal
  A}_{\theta}$ to 
${\tilde A}_i,{\tilde\lambda}\in{\cal A}_{\tilde\theta}$  which is
`determined' by gauge equivalence. This transformation has large
ambiguity due to path dependence in $\theta$-space. However this
ambiguity is negligible in particular in the $U(1)$ case and in rough
approximation of ignoring ${\cal O}(\partial\hF)$. This fact justifies
the equivalence of noncommutative Dirac-Born-Infeld Lagrangian
(\ref{eq:DBIhat}) in the $\theta$-space. 

However the ambiguity is no longer negligible in the $U(N)$ case or in
the $U(1)$ case if $\theta\ne 0$ and higher derivative correction is
considered because the path dependence (\ref{eq:del12A}) is not of the
form of gauge transformation. So if  one considers higher derivative
correction from the Dirac-Born-Infeld action or the $U(N)$
generalization of (\ref{eq:DBIhat}) by using transformation determined
only by gauge equivalence, we need a more careful argument.
Geometrical interpretation of the variation with respect to $\theta$
such as (\ref{eq:delta}) would be required.  

\section*{{\it Note Added}}

Eq.(\ref{eq:del12A}) can be rewritten as follows:\footnote{
Eq.(\ref{eq:del12Fu1}) can also be rewritten as follows:
\begin{eqnarray}
&&[\delta_1,\delta_2]\hF_{ij}
={1\over 16}\dt_2^{kl}\dt_1^{pq}\biggr(4(i[\hD_p\hF_{ik},\hD_q\hF_{jl}]-i[\hD_k\hF_{ip},\hD_l\hF_{jq}]+[[\hF_{ik},\hF_{jp}]+[\hF_{ip},\hF_{jk}],\hF_{lq}])\nonumber\\
&&+2i[\hF_{kp},\hD_l\hD_q\hF_{ij}+\hD_q\hD_l\hF_{ij}]
-i[\hF_{ij},\mfrac12\{\hA_k,\{\hA_p,\hF_{lq}\}\}+\mfrac12\{\hA_p,\{\hA_k,\hF_{lq}\}\}\nonumber\\
&&+\mfrac12[[\hA_k,\hA_p],\partial_l\hA_q+\partial_q\hA_l]
-i[\partial_p\hA_k,\partial_l\hA_q]+i[\partial_k\hA_p,\partial_q\hA_l]]\biggr)
+({\rm \alpha,\beta\ dependent\ terms}).
\end{eqnarray}
}

\begin{eqnarray}
[\delta_1,\delta_2]A_i
&=&{1\over 16}\dt_2^{kl}\dt_1^{pq}\biggr( 2i[\hF_{kp},\hD_l\hF_{qi}+\hD_q\hF_{li}]\nonumber\\
&&+\hD_i\big(\mfrac12\{\hA_k,\{\hA_p,\hF_{lq}\}\}+\mfrac12\{\hA_p,\{\hA_k,\hF_{lq}\}\}\nonumber\\
&&+\mfrac12[[\hA_k,\hA_p],\partial_l\hA_q+\partial_q\hA_l]-i[\partial_p\hA_k,\partial_l\hA_q]+i[\partial_k\hA_p,\partial_q\hA_l]\big)\biggr)\nonumber\\
&&+\hD_i({\rm \alpha,\beta\ dependent\ terms}).
\end{eqnarray}
The first term on the right hand side is `local' and
gauge-covariant. So, it can be absorbed by a {\it field redefinition}
of the gauge field ${\hA_i}$.\footnote{
This was suggested by Y.~Okawa and E.~Witten.}
The rest is of the form of gauge transformation.
The former corresponds to the ambiguity of ${\hat G}_i^{(n)F}$ and the
latter to that of ${\hat G}_i^{(n)}$ in (\ref{eq:A0n}).

This shows that, if we require a `physical input' such that a noncommutative
gauge field would be defined only up to {\it field redefinitions},
physics does not generally depend on paths in
 $\theta$-space.\footnote{
Of course, by a choice of a path (or a field redefinition), the
functional form of the action changes.}

\section*{Acknowledgments}
We would like to thank K.~Hashimoto, H.~Hata and T.~Kawano for
valuable discussions and comments. We appreciate hospitality of the
organizers of Summer Institute '99 where a part of this work was
discussed.
%\footnote{
%%WinTpicVersion2.16
%\unitlength 0.1in
%\begin{picture}(31.90,30.60)(20.00,-34.10)
%% POLYGON 2 0 3 0
%% 6 2000 3810 5190 3800 4140 1120 2650 750 2650 750 2000 3810
%% 
%\special{pn 8}%
%\special{pa 2000 3410}%
%\special{pa 5190 3400}%
%\special{pa 4140 720}%
%\special{pa 2650 350}%
%\special{pa 2650 350}%
%\special{pa 2000 3410}%
%\special{fp}%
%% CIRCLE 2 0 3 0
%% 4 2010 3810 2240 3810 2240 3810 2060 3530
%% 
%\special{pn 8}%
%\special{ar 2010 3410 230 230  4.8890978 6.2831853}%
%% CIRCLE 2 0 3 0
%% 4 5190 3800 5110 3590 5110 3600 4990 3800
%% 
%\special{pn 8}%
%\special{ar 5190 3400 225 225  3.1415927 4.3318826}%
%% CIRCLE 2 0 3 0
%% 4 2660 760 2840 940 2840 940 2920 810
%% 
%\special{pn 8}%
%\special{ar 2660 360 255 255  0.1899883 0.7853982}%
%% STR 2 0 3 0
%% 3 2350 3600 2350 3700 2 0
%% 50
%\put(23.5000,-33.0000){\makebox(0,0)[lb]{50}}%
%% STR 2 0 3 0
%% 3 2200 3250 2200 3350 2 0
%% 30
%\put(22.0000,-29.5000){\makebox(0,0)[lb]{30}}%
%% STR 2 0 3 0
%% 3 4550 3610 4550 3710 2 0
%% 60
%\put(45.5000,-33.1000){\makebox(0,0)[lb]{60}}%
%% STR 2 0 3 0
%% 3 4510 2890 4510 2990 2 0
%% 20
%\put(45.1000,-25.9000){\makebox(0,0)[lb]{20}}%
%% STR 2 0 3 0
%% 3 3010 910 3010 1010 2 0
%% x
%\put(30.1000,-6.1000){\makebox(0,0)[lb]{x}}%
%% LINE 2 1 3 0
%% 4 2660 750 5180 3800 4150 1130 2000 3810
%% 
%\special{pn 8}%
%\special{pa 2660 350}%
%\special{pa 5180 3400}%
%\special{da 0.070}%
%\special{pa 4150 730}%
%\special{pa 2000 3410}%
%\special{da 0.070}%
%\end{picture}%
%}
T.~A.\ and I.~K.\ are
supported in part by the Grant-in-Aid (\#04319) and (\#9858),
respectively, from the Ministry of Education, Science, Sports and
Culture.

We would also like to thank Y.~Okawa and E.~Witten for reading the
first version of this paper and giving us valuable comments.

\end{document}